\documentclass[12pt]{article}
\usepackage{amsmath}
\usepackage{amsfonts}
\usepackage{amssymb}
\usepackage{latexsym}
\usepackage{graphicx}
\usepackage[english]{babel}
\input epsf.sty

\begin{document}
{}~ {}~ \hfill\vbox{\hbox{USTC-ICTS-07-02}}\break

\vskip .4cm \centerline{\Large \bf Probing $\alpha$-Vacua of Black Holes in LHC} \vskip .6cm
\medskip

\vspace*{4.0ex} \centerline{\large \rm Tower Wang} \vspace*{4.0ex}

\centerline{\it Institute of Theoretical Physics, Chinese Academy of Sciences,}
\centerline{\it P. O. Box 2735£¬ Beijing 100080, China}
\centerline{\it Interdisciplinary Center for Theoretical Study,}
\centerline{\it University of Science and Technology of China,}
\centerline{\it Hefei, Anhui 230026, China}

\centerline{wangtao218@itp.ac.cn}

\vspace*{5.0ex} \centerline{\bf Abstract}\bigskip

Motivated by the idea of $\alpha$-vacua in Schwarzschild spacetime,
we studied the deformed spectrum of Hawking radiation. Such a
deformation would leave signatures on the small black hole
evaporation in LHC because their vacuum deviates from the Unruh
state.

\vfill \eject

\baselineskip=18pt

For massive scalar fields there is a family of de Sitter-invariant
states, including the Hartle-Hawking vacuum state as a special case
\cite{Mottola85,Allen85}. These states are well known as the
so-called $\alpha$-vacua because they are related to Hartle-Hawking
vacuum by a new parameter $\alpha$ \cite{Allen85}.\footnote{Actually
the $\alpha$-vacua in \cite{Allen85} are defined with two parameters
$\alpha$ and $\beta$. We will use $\gamma$ instead of $\beta$
following the notations of \cite{CM0610}.} The choice of physical
vacuum from these states might be imprinted on the spectrum of CMBR
(see e.g. \cite{D0205}). The application of $\alpha$-vacua to
cosmological dark energy can be found in \cite{AMM0612} as a recent
review.

Starting with Hartle-Hawking vacuum, Chamblin and Michelson in
\cite{CM0610} constructed the $\alpha$-vacua of scalar field in
Schwarzschild, Schwarzschild-dS and Schwarzschild-AdS black hole
spacetimes. However, it is generally conceived that at late times, a
black hole formed from gravitational collapse is well-approximated
by an eternal black hole with the scalar field in the Unruh state
\cite{Unruh76}, rather than in the Hartle-Hawking state
\cite{HH76}. On the other hand, assuming large extra dimensions
and TeV Plank scale, small black holes would be produced and decay
in LHC \cite{BF9906,GT0106,DL0106}, thus provide the first
experimental test of Hawking's radiation hypothesis
\cite{Hawking75}. Their sudden evaporation takes place not very long
after the formation, so naturally their vacuum deviates from the
Unruh state. Fortunately, since the discussion in \cite{CM0610} does
not depend heavily on the details of vacuum with which one starts,
the $\alpha$-vacua from the Unruh state can also be constructed, in
hopes of characterizing such a deviation with new parameters.

In this paper, we explore the modified spectrum of Hawking radiation
for scalars, considering effects of the $\alpha$-vacua of
Schwarzschild spacetime. If small black holes are indeed produced in
LHC abundantly, such a modification may leave observable
fingerprints on their evaporation spectrum.

Before getting the spectrum, we would like to work over some details
about the $\alpha$-vacua. We first introduce a set of global
coordinates of Schwarzschild spacetime, by embedding it into
$\mathcal{M}^{6,1}$. The global coordinates are a byproduct of this
paper, in which the symmetry under antipodal map is more manifest
than in other coordinates. After partly solving the Klein-Gordon
equation, we then deduce a more exact expression of $\alpha$-vacua,
which relates to the usual Hartle-hawking vacuum modes via a trivial
(nonmixing) Bogoliubov transformation and a Mottola-Allen
transformation.

Subsequently, in order to search some signatures of $\alpha$-vacua
in LHC phenomenologically, we turn to the Unruh state and the
corresponding $\alpha$-vacua, and derive the evaporation spectrum
following the standard procedure, i.e., by picking up the Bogoliubov
coefficients. The spectrum depends on two new parameters $\alpha$
and $\gamma$. When $\alpha\neq0$, it deviates from ordinary greybody
spectrum of Hawking radiation. So at the end of this paper, we
discussed implications of our result for small black holes in large
extra dimensional scenarios.

In static coordinates, the Schwarzschild metric is
\begin{equation}
ds^2=-\left(1-\frac{2M}{r}\right)dt^2+\left(1-\frac{2M}{r}\right)^{-1}dr^2+r^2(d\theta^2+\sin^2\theta d\varphi^2).
\end{equation}
It can be embedded into $\mathcal{M}^{6,1}$
\begin{equation}
ds^2=-dX^2_0+dX^2_1+dX^2_2+dX^2_3+dX^2_4+dX^2_5+dX^2_6
\end{equation}
by setting \cite{Gibbons86,CM0610}
\begin{eqnarray}
\nonumber &&X_1=r\sin\theta\cos\varphi,~~~~X_2=r\sin\theta\sin\varphi,~~~~X_3=r\cos\theta,\\
\nonumber &&X_4=-2M\sqrt{\frac{2M}{r}}+4M \sqrt{\frac{r}{2M}},~~~~X_5=2\sqrt{3}M\sqrt{\frac{2M}{r}},\\
&&X_6=4M\sqrt{1-\frac{2M}{r}}\cosh\left(\frac{t}{4M}\right),~~~~X_0=4M\sqrt{1-\frac{2M}{r}}\sinh\left(\frac{t}{4M}\right).
\end{eqnarray}

The spacetime outside the horizon of Schwarzschild black hole is
given as the algebraic variety determined by the three polynomials
\cite{Gibbons86,CM0610}
\begin{eqnarray}
\nonumber \frac{4}{3}X_5^2+X_6^2-X_0^2&=&16M^2,\\
\nonumber X_5^4(X_1^2+X_2^2+X_3^2)&=&576M^6,\\
\sqrt{3}X_4X_5+X_5^2&=&24M^2.
\end{eqnarray}

The global coordinates are related to static coordinates via
\begin{equation}\label{id}
\left(e^{\frac{t}{2M}},r,\theta,\varphi\right)_{\mathrm{static}}\rightarrow\left(\frac{\cos\sigma+\tanh\tau}{\cos\sigma-\tanh\tau},\frac{2M}{\cosh^2\tau\sin^2\sigma},\theta,\varphi\right)_{\mathrm{global}},
\end{equation}
which is resulted from the definition
\begin{eqnarray}
\nonumber &&X_1=\frac{2M\sin\theta\cos\varphi}{\cosh^2\tau\sin^2\sigma},~~~~X_2=\frac{2M\sin\theta\sin\varphi}{\cosh^2\tau\sin^2\sigma},~~~~X_3=\frac{2M\cos\theta}{\cosh^2\tau\sin^2\sigma},\\
\nonumber &&X_4=-2M\cosh\tau\sin\sigma+\frac{4M}{\cosh\tau\sin\sigma},~~~~X_5=2\sqrt{3}M\cosh\tau\sin\sigma,\\
&&X_6=4M\cosh\tau\cos\sigma,~~~~X_0=4M\sinh\tau,
\end{eqnarray}
and gives the form of metric
\begin{eqnarray}\label{global}
\nonumber ds^2&=&\frac{16M^2}{\cosh^6\tau\sin^6\sigma}\{\sin^2\sigma[\sinh^2\tau(1+\cosh^2\tau\sin^2\sigma)-\cosh^4\tau\sin^4\sigma]d\tau^2\\
\nonumber &&+\cosh^2\tau[\cos^2\sigma(1+\cosh^2\tau\sin^2\sigma+\cosh^4\tau\sin^4\sigma)+\cosh^6\tau\sin^6\sigma]d\sigma^2\\
\nonumber &&+2\sinh\tau\cosh\tau\sin\sigma\cos\sigma(1+\cosh^2\tau\sin^2\sigma+\cosh^4\tau\sin^4\sigma)d\tau d\sigma\}\\
&&+\frac{4M^2}{\cosh^4\tau\sin^4\sigma}(d\theta^2+\sin^2\theta d\varphi^2).
\end{eqnarray}
Clearly the global coordinates presented above are quite similar to
but more complicated than those for de Sitter spacetime
\cite{SSV0110}.

Correspondingly the measure is
\begin{equation}
\sqrt{-g}=\left|\frac{64M^4\sin\theta}{\cosh^6\tau\sin^7\sigma}\right|,
\end{equation}
while the non-vanishing components of contravariant metric tensor
are
\begin{eqnarray}
\nonumber g^{\tau\tau}&=&-\frac{1}{16M^2}[\cos^2\sigma(1+\cosh^2\tau\sin^2\sigma+\cosh^4\tau\sin^4\sigma)+\cosh^6\tau\sin^6\sigma],\\
\nonumber g^{\tau\sigma}&=&g^{\sigma\tau}=\frac{\sinh\tau\sin\sigma\cos\sigma}{16M^2\cosh\tau}(1+\cosh^2\tau\sin^2\sigma+\cosh^4\tau\sin^4\sigma),\\
\nonumber g^{\sigma\sigma}&=&-\frac{\sin^2\sigma}{16M^2\cosh^2\tau}[\sinh^2\tau(1+\cosh^2\tau\sin^2\sigma)-\cosh^4\tau\sin^4\sigma],\\
g^{\theta\theta}&=&\frac{\cosh^4\tau\sin^4\sigma}{4M^2},~~~~g^{\varphi\varphi}=\frac{\cosh^4\tau\sin^4\sigma}{4M^2\sin^2\theta}.
\end{eqnarray}

The covariant Klein-Gordon equation
\begin{equation}\label{KG}
(\Box_x-\mu^2)\Phi(x)=\frac{1}{\sqrt{-g}}\partial_{\mu}(\sqrt{-g}g^{\mu\nu}\partial_{\nu}\Phi)-\mu^2\Phi=0
\end{equation}
in global coordinates is too complicated to be fully solved by brute
force. However, we can still get some details from it as follows.

In terms of global coordinates, the antipodal map
\cite{Allen85,CM0610} $x\rightarrow x_A$ reads
\begin{equation}
(\tau,\sigma,\theta,\varphi)\rightarrow(-\tau,\pi-\sigma,\pi-\theta,\varphi\pm\pi).
\end{equation}
It is easy to see the metric (\ref{global}) is manifestly invariant
under this antipodal transformation and further
\begin{eqnarray}\label{Hadamard1}
\nonumber &&(\Box_x-\mu^2)G_0^{(1)}(x,y)=0\\
\Rightarrow&&(\Box_{x_A}-\mu^2)G_0^{(1)}(x_A,y)=(\Box_x-\mu^2)G_0^{(1)}(x_A,y)=0
\end{eqnarray}
for the Hadamard function $G_0^{(1)}(x,y)$. That is to say, the
Hadamard functions $G_0^{(1)}(x,y)$ and $G_0^{(1)}(x_A,y)$ satisfy
the same equation. This property is most manifest in global
coordinates. We emphasize the property here and confirm the
existence of a complete set of orthonormal modes obeying
\begin{equation}\label{Amode1}
\phi_{\omega,l,m}(x_A)=\phi^{*}_{\omega,l,m}(x)
\end{equation}
in the following, because they play important roles in proving that
the $\alpha$-vacua respect symmetries of the spacetime
\cite{Allen85,CM0610}. For de Sitter spacetime, the symmetry is the
$O(1,4)$ group. While for Schwarzchild spacetime, it is the
``$CPT$'' invariance introduced in \cite{Gibbons86} and recounted in
\cite{CM0610}.

In \cite{CM0610}, it was argued that one can choose the modes
satisfying (\ref{Amode1}) because the equation of motion is
invariant under complex conjugation and under the antipodal map. It
seems to me this argument guarantees only the existence of certain
special solution of differential equation (\ref{KG}) which satisfies
$\Phi(x_A)=\Phi^{*}(x)$, rather than the existence of a set of
Hartle-Hawking vacuum modes $\phi_{\omega,l,m}$ satisfying
(\ref{Amode1}). We would like to fill this gap and make the argument
in \cite{CM0610} more solid, by precisely constructing these modes
of Hartle-Hawking vacuum.

Our construction will be based on some results presented in
\cite{FN}. In the book \cite{FN}, a complete set of Hartle-Hawking
modes $\hat{\phi}^{(1)}_{\omega,l,m}$,
$\hat{\phi}^{(2)}_{\omega,l,m}$, $\hat{\phi}^{(3)}_{\omega,l,m}$ and
$\hat{\phi}^{(4)}_{\omega,l,m}$ has been
constructed.\footnote{Different from chapter 11.2 of book \cite{FN},
we will use notations $\hat{\phi}^{(1)}_{\omega,l,m}$,
$\hat{\phi}^{(2)}_{\omega,l,m}$, $\hat{\phi}^{(3)}_{\omega,l,m}$ and
$\hat{\phi}^{(4)}_{\omega,l,m}$ instead of $\varphi_J^{d'}$,
$\varphi_J^{p'}$, $\varphi_J^{u'}$ and $\varphi_J^{n'}$ respectively
for self-consistent of this paper.} These modes are orthonormal and
meet conditions \cite{FN}
\begin{eqnarray}
\nonumber \hat{\phi}^{(1)}_{\omega,l,m}(-U,-V,\theta,\varphi)&=&\hat{\phi}^{(2)*}_{\omega,l,m}(U,V,\theta,\varphi),\\
\hat{\phi}^{(3)}_{\omega,l,m}(-U,-V,\theta,\varphi)&=&\hat{\phi}^{(4)*}_{\omega,l,m}(U,V,\theta,\varphi).
\end{eqnarray}
Here $U$, $V$ are Kruskal coordinates. In our global coordinates,
the conditions are translated into
\begin{eqnarray}\label{FNmode1}
\nonumber \hat{\phi}^{(1)*}_{\omega,l,m}(-\tau,\pi-\sigma,\theta,\varphi)&=&\hat{\phi}^{(2)}_{\omega,l,m}(\tau,\sigma,\theta,\varphi),\\
\hat{\phi}^{(3)*}_{\omega,l,m}(-\tau,\pi-\sigma,\theta,\varphi)&=&\hat{\phi}^{(4)}_{\omega,l,m}(\tau,\sigma,\theta,\varphi).
\end{eqnarray}
At the same time, it is trivial to show the factorization
\begin{equation}\label{factori}
\hat{\phi}^{(i)}_{\omega,l,m}(x)=\hat{f}^{(i)}_{\omega,l,m}(\tau,\sigma)Y_{l,m}(\theta,\varphi),~~~~i=1,2,3,4
\end{equation}
is valid utilizing equation (\ref{KG}). We stress that such a
factorization is valid in the whole spacetime. On boundaries
$\mathcal{H}^{\pm}$ especially, the book \cite{FN} gave analytic
expressions of $\hat{\phi}^{(i)}_{\omega,l,m}$, and one can show
they are factorized as (\ref{factori}) indeed. Combining
(\ref{FNmode1}), (\ref{factori}) and the following formula for
spherical harmonic functions
\begin{equation}
Y_{l,m}(\pi-\theta,\varphi\pm\pi)=(-1)^lY_{l,m}(\theta,\varphi)
\end{equation}
together, we can check the relations
\begin{eqnarray}
\nonumber \hat{\phi}^{(1)*}_{\omega,l,m}(x_A)&=&(-1)^l\hat{\phi}^{(2)}_{\omega,l,m}(x),\\
\hat{\phi}^{(3)*}_{\omega,l,m}(x_A)&=&(-1)^l\hat{\phi}^{(4)}_{\omega,l,m}(x),
\end{eqnarray}
and construct a set of new orthonormal modes $\phi_{\omega,l,m}$ by
the trivial Bogoliubov transformation \cite{Allen85}
\begin{eqnarray}\label{Amode2}
\nonumber \phi^{(1)}_{\omega,l,m}(x)&=&\frac{1}{\sqrt{2}}e^{i\pi l/2}[\hat{\phi}^{(1)}_{\omega,l,m}(x)+\hat{\phi}^{(2)}_{\omega,l,m}(x)],\\
\nonumber \phi^{(2)}_{\omega,l,m}(x)&=&\frac{1}{\sqrt{2}}e^{i\pi(l+1)/2}[\hat{\phi}^{(1)}_{\omega,l,m}(x)-\hat{\phi}^{(2)}_{\omega,l,m}(x)],\\
\nonumber \phi^{(3)}_{\omega,l,m}(x)&=&\frac{1}{\sqrt{2}}e^{i\pi l/2}[\hat{\phi}^{(3)}_{\omega,l,m}(x)+\hat{\phi}^{(4)}_{\omega,l,m}(x)],\\
\phi^{(4)}_{\omega,l,m}(x)&=&\frac{1}{\sqrt{2}}e^{i\pi(l+1)/2}[\hat{\phi}^{(3)}_{\omega,l,m}(x)-\hat{\phi}^{(4)}_{\omega,l,m}(x)].
\end{eqnarray}

The modes (\ref{Amode2}) form a complete set of orthonormal modes.
In particular,
\begin{enumerate}
\item $\phi^{(i)}_{\omega,l,m}(x_A)=\phi^{(i)*}_{\omega,l,m}(x)$.
\item $(\phi^{(i)}_{\omega,l,m},\phi^{(i')}_{\omega',l',m'})=\delta_{i,i'}\delta_{\omega,\omega'}\delta_{l,l'}\delta_{m,m'}$ and $(\phi^{(i)}_{\omega,l,m},\phi^{(i')*}_{\omega',l',m'})=0$.
\item The set of $\phi_{\omega,l,m}$ is complete and spans the space of $\hat{\phi}_{\omega,l,m}$.
\end{enumerate}

The $\alpha$-vacua are constructed from this set of modes by a
Mottola-Allen transformation \cite{Allen85,CM0610}
\begin{equation}\label{AM1}
\tilde{\phi}^{(i)}_{\omega,l,m}(x)=\cosh\alpha\phi^{(i)}_{\omega,l,m}(x)+e^{i\gamma}\sinh\alpha\phi^{(i)*}_{\omega,l,m}(x),~~~~\alpha\geq0,~~-\pi<\gamma\leq\pi.
\end{equation}
One should notice that $\hat{\phi}^{(i)}_{\omega,l,m}(x)$ and
$\phi^{(i)}_{\omega,l,m}(x)$ are of positive frequencies with
respect to the affine parameters on $\mathcal{H}^{\pm}$. However,
this is not true for $\tilde{\phi}^{(i)}_{\omega,l,m}(x)$ since the
transformation (\ref{AM1}) mixes modes of the same frequency but
with different sign. Or equivalently, from another point of view
based on (\ref{Amode1}), it mixes modes on the antipodal points $x$
and $x_A$. The modes (\ref{AM1}) are taken as a new ``vacuum'' state
\cite{Mottola85,Allen85} because this transformation is (the
Bogoliubov coefficients are) frequency independent and preserves
orthonormality.

The scalar field in (\ref{KG}) may be decomposed in different bases
if we consider different vacua,
\begin{eqnarray}\label{Amode2}
\nonumber \Phi(x)&=&\sum_{i,\omega,l,m}\left[a^{(i)}_{\omega,l,m}\phi^{(i)}_{\omega,l,m}(x)+a^{(i)\dag}_{\omega,l,m}\phi^{(i)*}_{\omega,l,m}(x)\right]\\
&=&\sum_{i,\omega,l,m}\left[\tilde{a}^{(i)}_{\omega,l,m}\tilde{\phi}^{(i)}_{\omega,l,m}(x)+\tilde{a}^{(i)\dag}_{\omega,l,m}\tilde{\phi}^{(i)*}_{\omega,l,m}(x)\right].
\end{eqnarray}

The properties (\ref{Hadamard1}) and (\ref{Amode1}) are the major
tricks to study properties of two-point functions, and to prove that
$\alpha$-vacua respect the symmetries of the spacetime
\cite{Allen85,CM0610}. For example, equation (\ref{Amode1}) leads to
a relation between Hadamard function $G_{\alpha\gamma}^{(1)}(x,y)$
for $\alpha$-vacua and $G_0^{(1)}(x,y)$, $D_0(x,y)$ for
Hartle-Hawking vacuum,
\begin{equation}\label{Hadamard2}
G_{\alpha\gamma}^{(1)}(x,y)=\cosh(2\alpha)G_0^{(1)}(x,y)+\cos\gamma\sinh(2\alpha)G_0^{(1)}(x_A,y)-\sin\gamma\sinh(2\alpha)D_0(x_A,y).
\end{equation}
From (\ref{Amode1}) it is clear that $G_0^{(1)}(x,y)$ and
$G_0^{(1)}(x_A,y)$ obey the same equation of motion and hence
respect symmetries of the spacetime. Therefore,
$G_{\alpha\gamma}^{(1)}(x,y)$ is $CPT$ \cite{Gibbons86,CM0610}
invariant for $\alpha$-vacua with $\gamma=0,\pi$. As explained in
\cite{Allen85} and reiterated in \cite{CM0610}, the $\alpha$-vacua
with $\sin\gamma\neq0$ break the time-reversal symmetry, to which we
will come back later when discussing small black holes in LHC.

In the previous part we focused on two tasks:
\begin{enumerate}
\item making it more manifest that $G_0^{(1)}(x_A,y)$ obeys the same equation as that of
$G_0^{(1)}(x,y)$, thus preserves the symmetries of the Schwarzschild
spacetime; and
\item establishing a complete set of orthonormal modes satisfying (\ref{Amode1}).
\end{enumerate}
Global coordinates (\ref{global}) of Scharzschild spacetime were a
byproduct during our research. The coordinates may be not necessary
here, but facilitate the first task in a way. Their physical
implications and applications in various aspects of Schwarzschild
black hole remain unclear, which we would like to study elsewhere in
the future.

In the following, to be brief, we will work in the matrix formalism.
That is, we will write a basis of modes in a column matrix and
multiply it by a square matrix to represent the Bogoliubov
transformation. For example, the trivial Bogoliubov transformation
(\ref{Amode2}) will be written concisely,
\begin{equation}
\phi_i=P_{ij}\hat{\phi}_j,
\end{equation}
where the collective subscript $i$ or $j$ denotes the complete set
of quantum numbers $\omega$,$l$,$m$ and superscripts ($i$) that must
be specified to describe a mode. Or even more compactly,
\begin{equation}
\phi=P\hat{\phi}.
\end{equation}
Likewise, the Mottola-Allen transformation (\ref{AM1}) will be
denoted in the form
\begin{equation}
\tilde{\phi}=\cosh\alpha\phi+e^{i\gamma}\sinh\alpha\phi^{*},~~~~\alpha\geq0,~~-\pi<\gamma\leq\pi.
\end{equation}

In the above, we have been dealing with the Hartle-Hawking type
$\alpha$-vacua, and focusing on some theoretical problems. In the
following, we will turn to a relatively independent issue -- a
phenomenological problem: calculating the evaporation spectrum for
the Unruh vacuum and those for its $\alpha$-vacua correspondingly.

For Unruh vacuum, the procedure is standard \cite{Unruh76,T9707}. If
we use $\psi'_{i}$ to label the positive frequency modes on
$\Im^{+}$ and $\psi_{i}$ to label the positive frequency modes on
$\Im^{-}$, then the Bogoliubov transformation
$\psi_{i}\rightarrow\psi'_{i}$, i.e., $a_{i}\rightarrow a'_{i}$ can
be written as
\begin{equation}\label{Bogo1}
\psi'=A\psi+B\psi^{*},
\end{equation}
or namely
\begin{equation}
a=a'A+a'^{\dag}B^{*},
\end{equation}
in which $A$ and $B$ are Bogoliubov transformation matrices, while
$a$ and $a'$ are row matrices with elements $a_{i}$ and $a'_{i}$
respectively. A solution of the field equation (\ref{KG}) can be
expanded as
\begin{equation}
\Phi(x)=\sum_{i}(a_{i}\psi_{i}+a^{\dag}_{i}\psi^{*}_{i})=\sum_{i}(a'_{i}\psi'_{i}+a_{i}'^{\dag}\psi'^{*}_{i}).
\end{equation}
It has been proved in \cite{T9707} that there is a relation between
the Bogoliubov coefficients
\begin{equation}\label{corollary}
B_{ij}=-e^{-\omega_{i}/(2T_{H})}A_{ij},
\end{equation}
and the late time particle flux through $\Im^{+}$ given a vacuum on
$\Im^{-}$ is determined by \cite{T9707}
\begin{equation}\label{BBd}
\left(BB^{\dag}\right)^{*}_{ii}=\frac{1}{e^{\omega_i/T_{H}}-1}.
\end{equation}
In (4+n)-dimensions \cite{MP86}, the Hawking temperature can be
traded for the black hole radius \cite{DL0106},
\begin{equation}\label{temperature}
T_{H}=\frac{n+1}{4\pi r_{H}}.
\end{equation}
Taking into consideration of greybody factor, the spectrum of energy
flux has the form
\begin{equation}
\frac{dE(\omega)}{dt}=\sum_l\sigma_{l,n}(\omega)\frac{\omega}{e^{\omega/T_{H}}-1}\frac{dk^{n+3}}{(2\pi)^{n+3}}.
\end{equation}

At low energy $\omega r_H\ll1$, an analytic expression for greybody
factor $\sigma_{l,n}$ has been derived in \cite{KM0203,K0402}. In
large extra dimensional scenarios, small black holes may emit scalar
fields in the bulk as well as on the brane. Our attention in this
paper will be ``localized'' on the brane with the help of analytic
formulas given in \cite{KM0203,K0402}, although the calculation in
the bulk can be accomplished in a similar way. In the massless
particle approximation, corresponding to Unruh vacuum (\ref{BBd}),
the low energy scalar spectrum on the brane is \cite{KM0203,K0402}
\begin{equation}\label{enfluxU}
\frac{d^2E(\omega)}{d\omega
dt}=r_H^{-1}\sum_l\frac{(2l+1)\Gamma(\frac{l+1}{n+1})^2\Gamma(1+\frac{l}{n+1})^2}{(n+1)^2\Gamma(\frac{1}{2}+l)^2\Gamma(1+\frac{2l+1}{n+1})^2}\left(\frac{\omega
r_H}{2}\right)^{2l}\frac{2(\omega r_H)^3}{e^{4\pi\omega
r_H/(n+1)}-1}.
\end{equation}

For the $\alpha$-vacua constructed from the Unruh state, we should
consider the following series of transformations
\begin{eqnarray}
\nonumber &\tilde{\psi}_{i}\rightarrow\psi_{i}\rightarrow\psi'_{i},&\\
&\tilde{a}_{i}\rightarrow a_{i}\rightarrow a'_{i}.&
\end{eqnarray}
In other words, we take the $\alpha$ state
\begin{equation}\label{AM2}
\tilde{\psi}=\cosh\alpha\psi+e^{i\gamma}\sinh\alpha\psi^{*},~~~~\alpha\geq0,~~-\pi<\gamma\leq\pi
\end{equation}
as the physical vacuum on $\Im^{-}$. Inversion of (\ref{AM2}) leads
to
\begin{equation}\label{AM3}
\psi=\cosh\alpha\tilde{\psi}-e^{i\gamma}\sinh\alpha\tilde{\psi}^{*}.
\end{equation}
At the same time, one can also formally write
\begin{equation}\label{Bogo2}
\psi'=\tilde{A}\tilde{\psi}+\tilde{B}\tilde{\psi}^{*},~~~~\tilde{\psi}=\tilde{A}'\psi'+\tilde{B}'\psi'^{*}.
\end{equation}

The expected number of particles in the $i$th mode is related to
$\left(\tilde{B}\tilde{B}^{\dag}\right)^{*}_{ii}$. From the
relations (\ref{Bogo1}), (\ref{AM3}) and (\ref{Bogo2}), one
immediately gets
\begin{equation}
\psi'=(A\cosh\alpha-Be^{-i\gamma}\sinh\alpha)\tilde{\psi}+(B\cosh\alpha-Ae^{i\gamma}\sinh\alpha)\tilde{\psi}^{*},
\end{equation}
thus
\begin{equation}
\tilde{B}=B\cosh\alpha-Ae^{i\gamma}\sinh\alpha.
\end{equation}
Multiplied by
\begin{equation}
\tilde{B}^{\dag}=B^{\dag}\cosh\alpha-A^{\dag}e^{-i\gamma}\sinh\alpha,
\end{equation}
it gives
\begin{equation}\label{tBtBdeg}
\tilde{B}\tilde{B}^{\dag}=\cosh^2\alpha\left(BB^{\dag}\right)+\sinh^2\alpha\left(AA^{\dag}\right)-\sinh\alpha\cosh\alpha[e^{i\gamma}\left(AB^{\dag}\right)+e^{-i\gamma}\left(BA^{\dag}\right)].
\end{equation}
As a consistency check, when $\alpha=0$, apparently it reduces to
the expected form $\tilde{B}\tilde{B}^{\dag}|_{\alpha=0}=BB^{\dag}$.
In virtue of (\ref{corollary}), (\ref{BBd}) and (\ref{temperature}),
one can simplify the diagonal entries in (\ref{tBtBdeg}) and write
down
\begin{equation}\label{BBdalpha}
\left(\tilde{B}\tilde{B}^{\dag}\right)^{*}_{ii}=\frac{\sinh^2\alpha
e^{4\pi\omega_ir_H/(n+1)}+\cos\gamma\sinh(2\alpha)e^{2\pi\omega_ir_H/(n+1)}+\cosh^2\alpha}{e^{4\pi\omega_ir_H/(n+1)}-1}.
\end{equation}

The absorption amplitude and greybody factor are caused by the
traversal of emitted particles in the gravitational background. They
are independent of the initial conditions, i.e., independent of the
vacuum we choose on $\Im^{-}$. As a result, given an $\alpha$ state
as the vacuum on $\Im^{-}$, at the late time, the low energy flux
through $\Im^{+}$ is
\begin{eqnarray}\label{enflux}
\nonumber \frac{d^2E(\omega)}{d\omega dt}&=&r_H^{-1}\sum_l\frac{(2l+1)\Gamma(\frac{l+1}{n+1})^2\Gamma(1+\frac{l}{n+1})^2}{(n+1)^2\Gamma(\frac{1}{2}+l)^2\Gamma(1+\frac{2l+1}{n+1})^2}\left(\frac{\omega r_H}{2}\right)^{2l}2(\omega r_H)^3\\
&&\times\frac{\sinh^2\alpha e^{4\pi\omega r_H/(n+1)}+\cos\gamma\sinh(2\alpha)e^{2\pi\omega r_H/(n+1)}+\cosh^2\alpha}{e^{4\pi\omega r_H/(n+1)}-1}.
\end{eqnarray}

For high energy emissions $\omega r_H\gg1$, equation (\ref{enflux})
predicts a divergent energy flux. This suggests the breakdown of
(\ref{BBdalpha}) and (\ref{enflux}) at very high energy. Indeed,
during derivation of the spectrum, we have neglected backreactions
to the black hole, which are supposed to be small at low energy. In
the high energy region, especially for small black holes, the
backreaction effect of the particle emission will be large, so the
spectrum (\ref{BBdalpha}) together with (\ref{enflux}) cannot be
trusted there any more.

Assuming large extra dimensions and TeV Plank scale, in
\cite{BF9906,GT0106,DL0106}, it has been proposed that small black
holes of TeV scale masses would be produced in LHC and provide the
first experimental test of Hawking's radiation hypothesis. On the
one hand, the lifetime of the small black holes in this scenario is
of order
\begin{equation}
\tau\sim
M_p^{*-1}\left(\frac{M}{M_p^{*}}\right)^{\frac{n+3}{n+1}}\sim r_H\left(\frac{M}{M_p^{*}}\right)^{\frac{n+2}{n+1}}
\end{equation}
or typically $10^{-26}$ second \cite{GT0106,K0402}, much shorter
than that of ordinary black holes in astrophysics. On the other
hand, for a black hole formed by collapse, at a sufficiently long
time after its formation, the Unruh state serves as a good boundary
condition of Green's function. While not long after its formation,
the boundary condition depends on details of the collapse. If we
introduce the $\alpha$-vacua (\ref{AM2}) as a new boundary
condition, there are two additional parameters $\alpha$ and
$\gamma$, which would capture some universal features of the black
hole formation in LHC and characterize the deviation of the vacuum
from the Unruh vacuum. In this sense, the physical values of
$\alpha$ and $\gamma$ cannot be determined
theoretically.\footnote{Of course, there might be some arguments
favoring vanishing values of $\alpha$ and $\gamma$. But that is a
matter subject to debate.} However, we observe that for non-eternal
black holes, possibly the value of $\alpha$ depends on their
lifetime. Specifically, we guess that $\alpha$ increases with
respect to the ratio $\frac{M_p^{*}}{M}$, and vanishes in the limit
$\frac{M_p^{*}}{M}\rightarrow0$. In large extra dimensional
scenarios, for small black holes produced in LHC,\footnote{Say,
typically we have $M_p^{*}\sim1$ TeV and $M\sim5$ TeV.} the ratio is
of order $1<\frac{M_p^{*}}{M}\lesssim10$, thus we can take $\alpha$
as a parameter of constant. If $\alpha$ is large enough, in the low
energy region $\omega r_H\ll1$, the greybody profile of Hawking
radiation will be deformed according to (\ref{enflux}).
Schwarzschild phase is an important stage during the evaporation of
small black holes. If a non-vanishing value of $\alpha$ is indeed
physical for these black holes, signatures of (\ref{enflux}) must be
imprinted on the evaporation spectrum, thus can be found out by a
detailed study of small black hole decay.

\begin{figure}
\center{\includegraphics[width=0.45\textwidth]{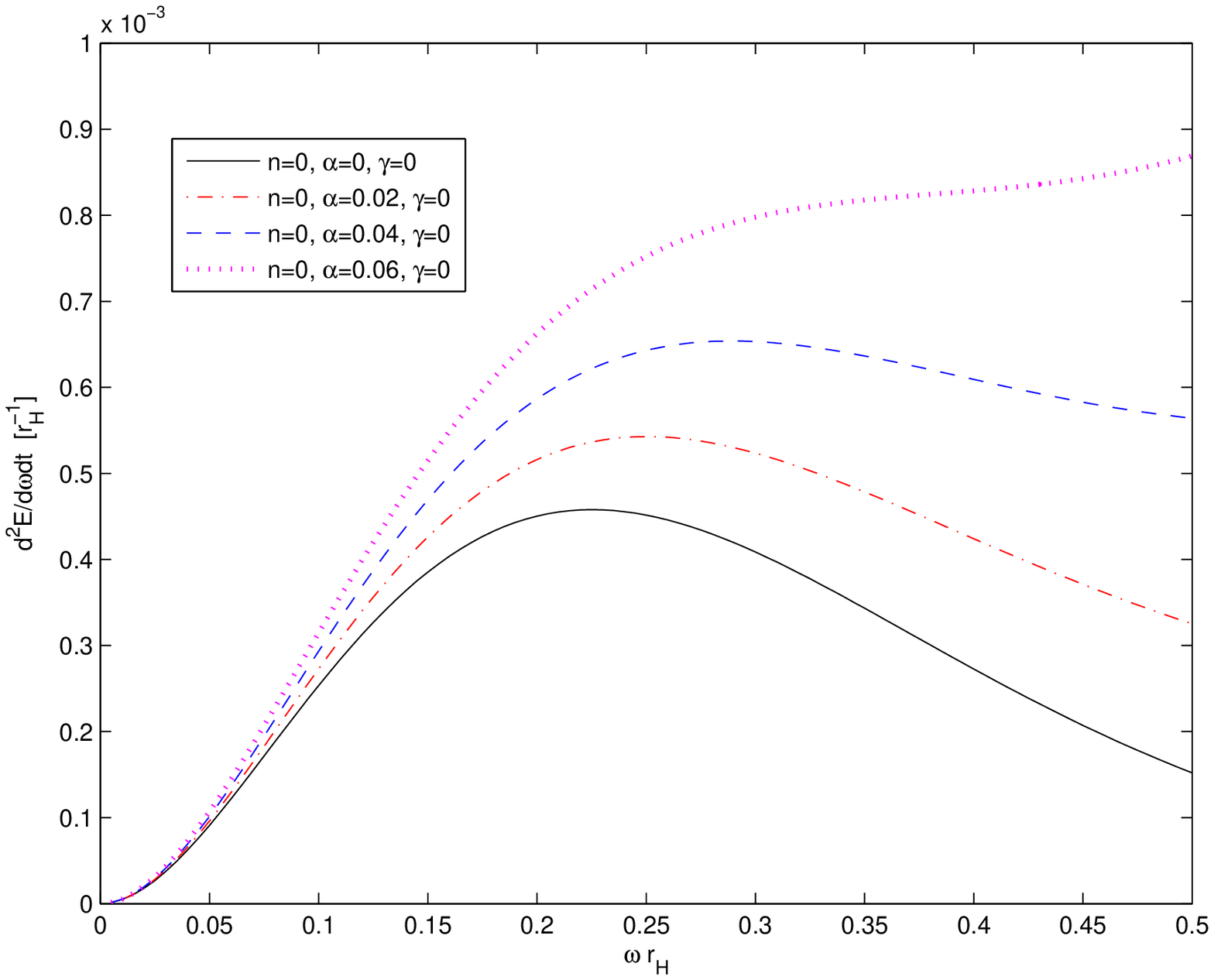}
\includegraphics[width=0.45\textwidth]{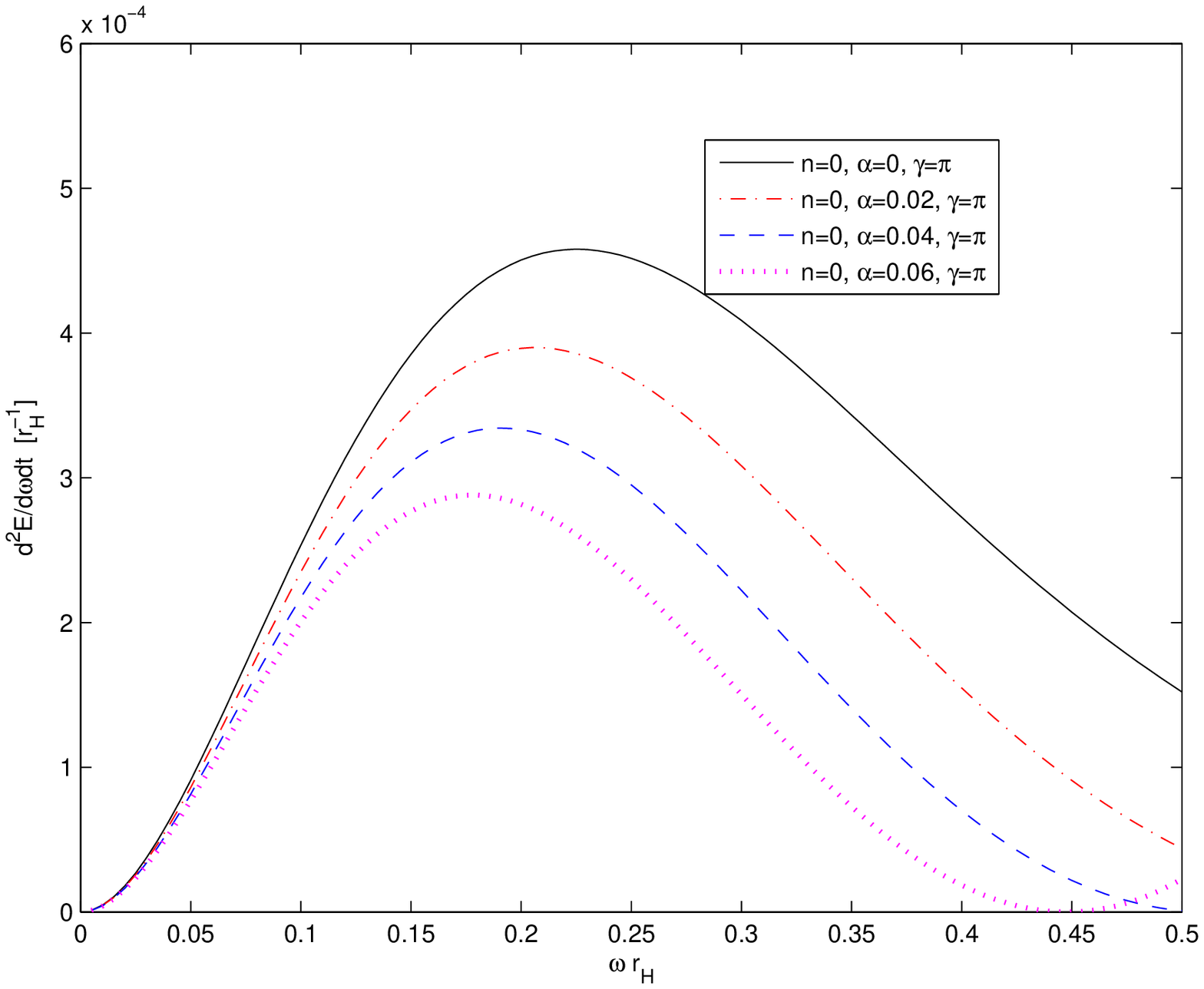}}\\
\caption{\textbf{The scalar energy flux $\frac{d^2E(\omega)}{d\omega
dt}$ on the brane as a function of $\omega r_H$ with  $n=0$,
$\alpha=0$ (black solid lines), $0.02$ (red dash-dotted lines),
$0.04$ (blue dashed lines), $0.06$ (magenta dotted lines) and
$\gamma=0$ (the left graph), $\pi$ (the right graph).}}\label{0}
\end{figure}

\begin{figure}
\center{\includegraphics[width=0.45\textwidth]{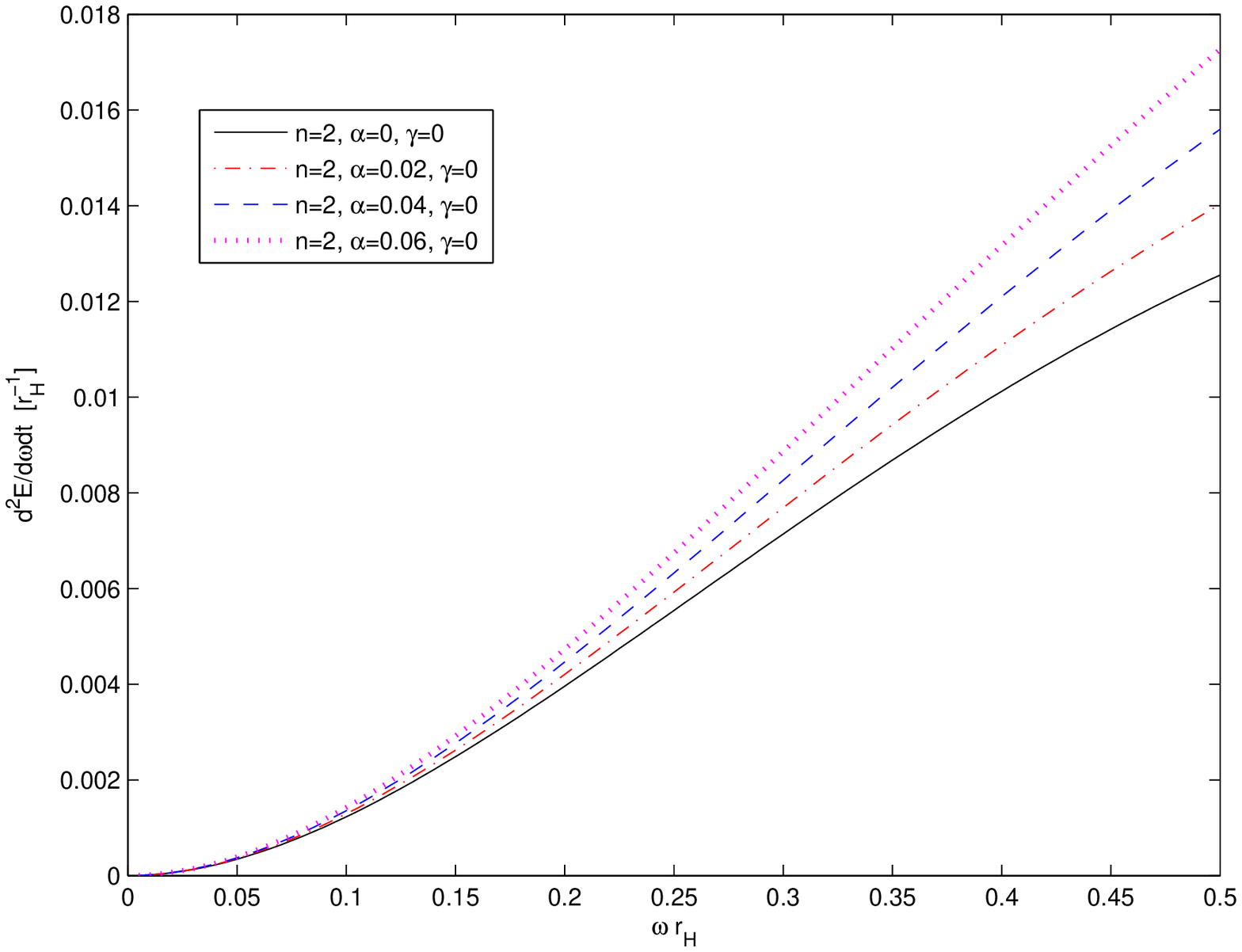}\includegraphics[width=0.45\textwidth]{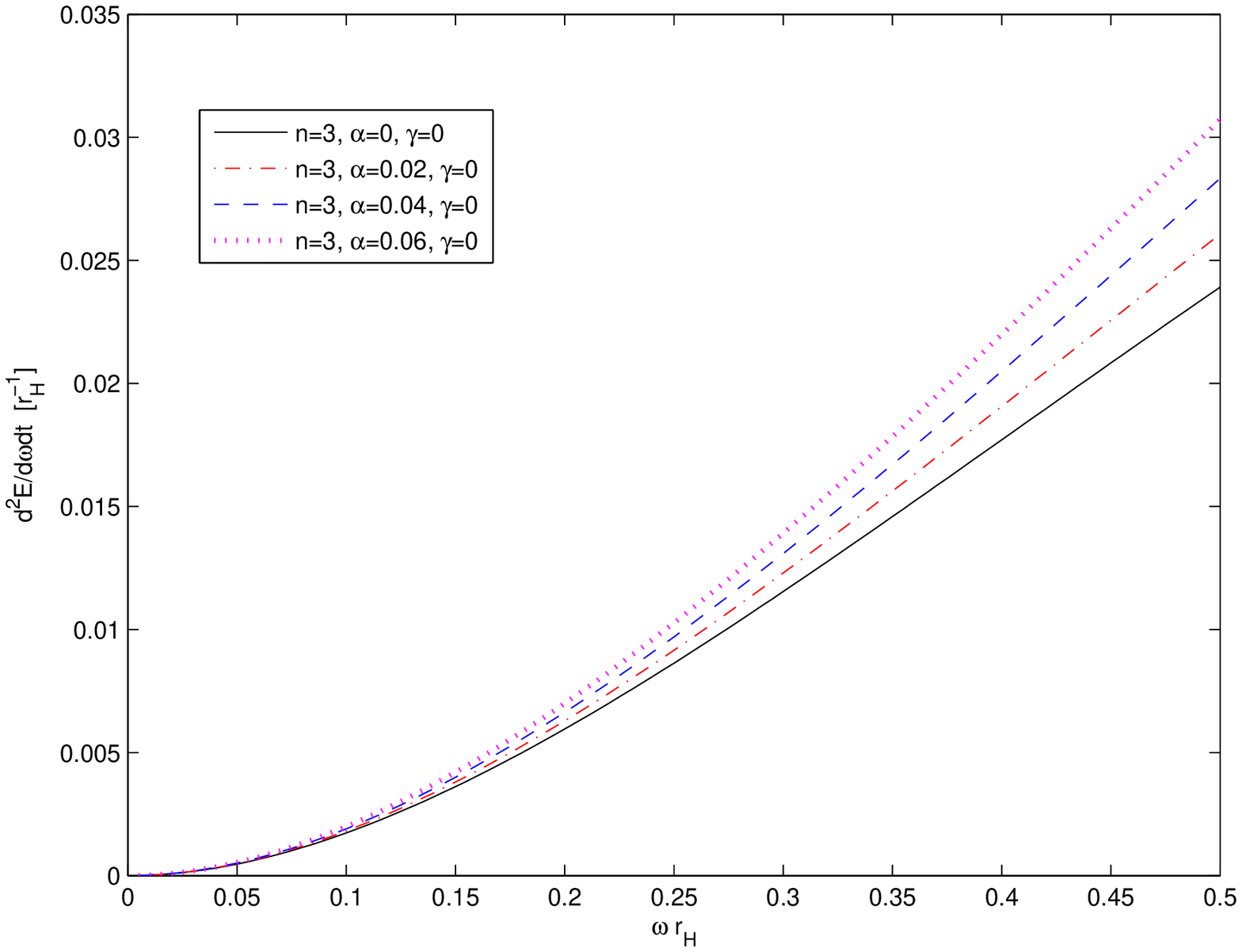}\\
\includegraphics[width=0.45\textwidth]{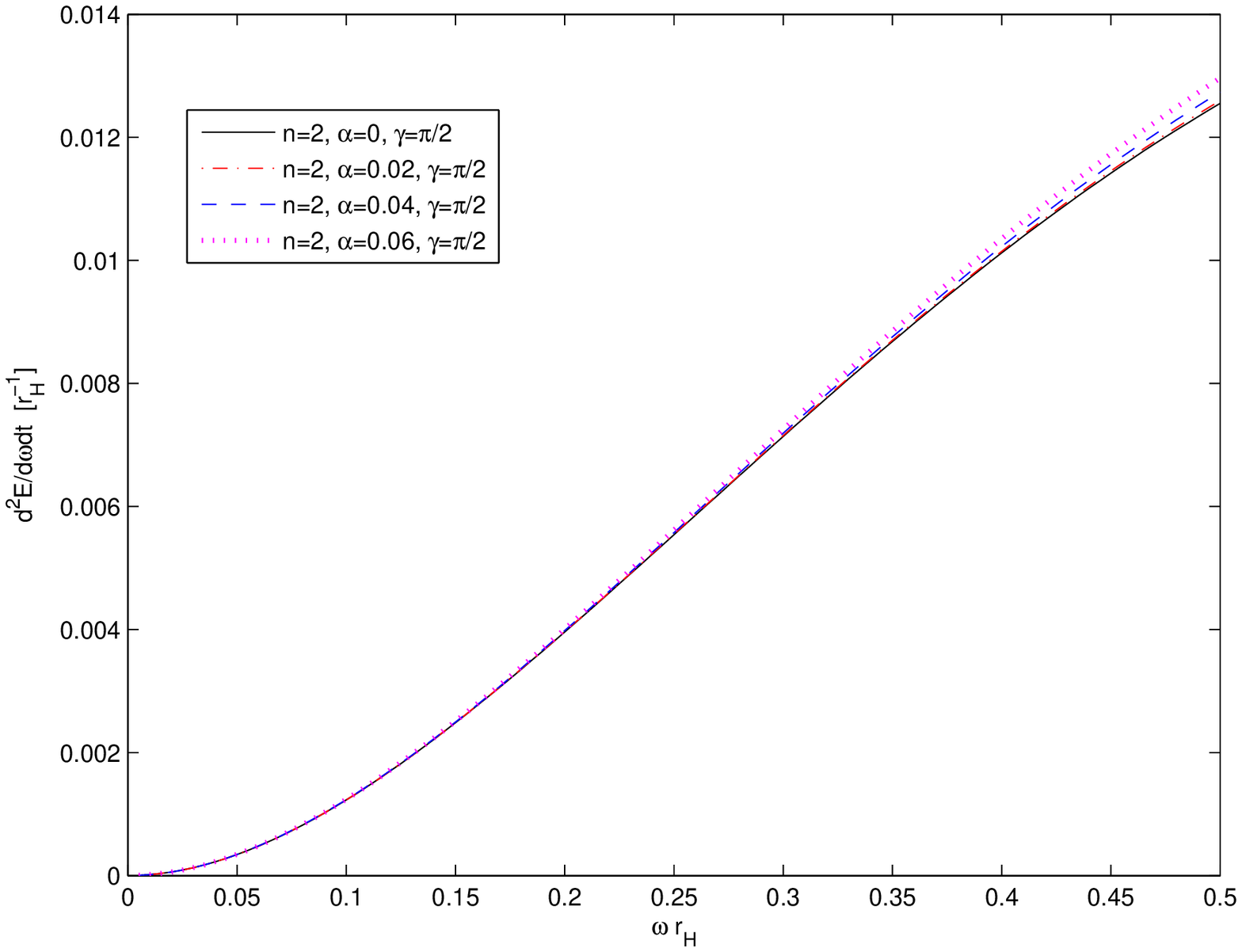}\includegraphics[width=0.45\textwidth]{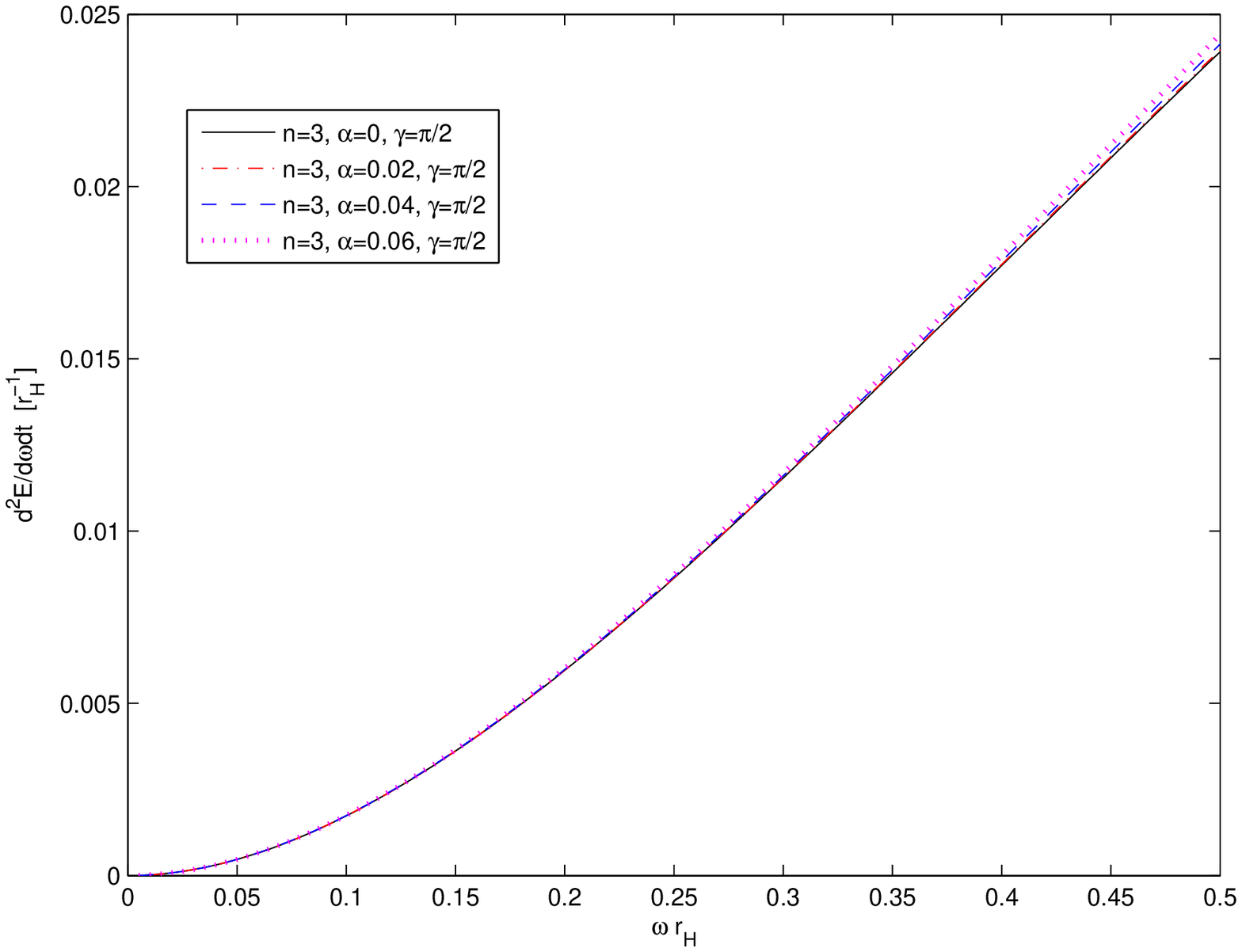}\\
\includegraphics[width=0.45\textwidth]{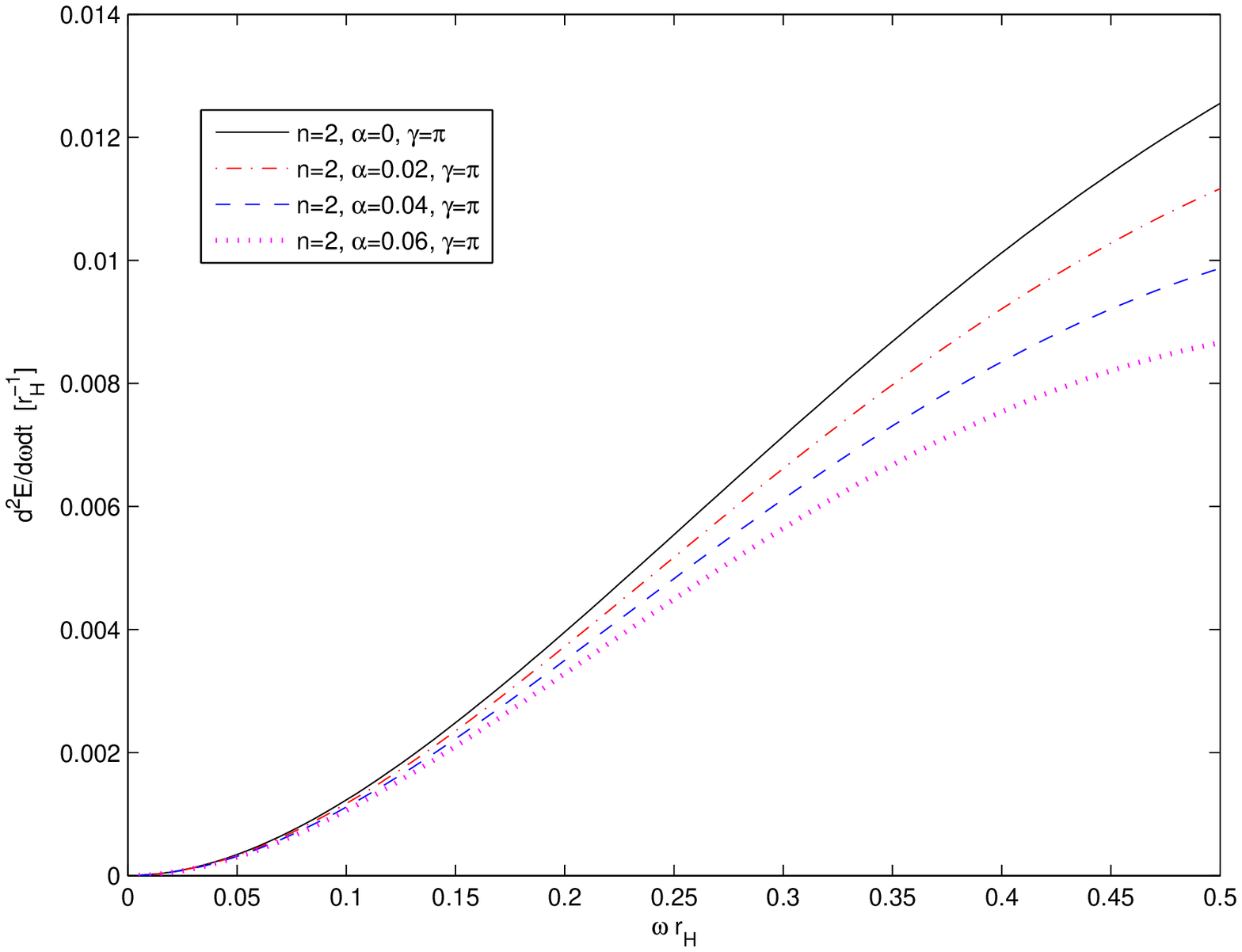}\includegraphics[width=0.45\textwidth]{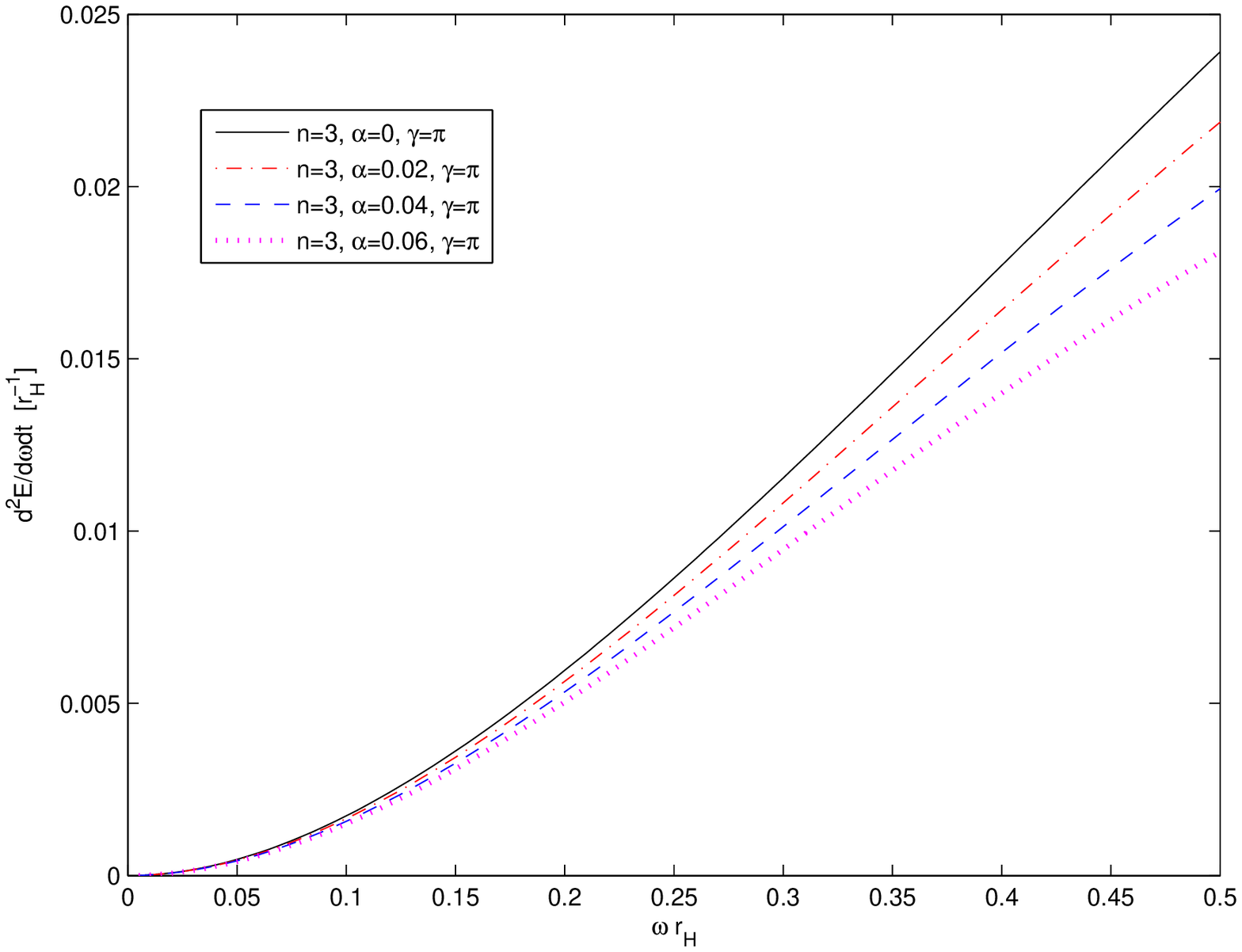}}\\
\caption{\textbf{The scalar energy flux $\frac{d^2E(\omega)}{d\omega
dt}$ on the brane as a function of $\omega r_H$ with $n=2$ (the left
graphs), $n=3$ (the right graphs), $\alpha=0$ (black solid lines),
$0.02$ (red dash-dotted lines), $0.04$ (blue dashed lines), $0.06$
(magenta dotted lines) and $\gamma=0$ (the upper graphs), $\pi/2$
(the middle graphs), $\pi$ (the lower graphs).}}\label{23}
\end{figure}

To give a dramatic impression, we show the low energy spectra of
scalar fluxes on the brane for various values of $n$, $\alpha$ and
$\gamma$ in figures \ref{0} and \ref{23}. When plotting the spectrum
according to (\ref{enflux}), we summed over $l$ up to the third
partial wave, since the contribution from higher partial waves is
negligible. In all of the figures, $\alpha=$0, 0.02, 0.04, 0.06 are
indicated by black solid lines, red dash-dotted lines, blue dashed
lines, and magenta dotted lines respectively. For comparison, we
show the low energy spectra without a large extra-dimension ($n=0$)
in figure \ref{0}. For scenarios with large extra-dimensions ($n=$2,
3), the spectra are depicted in figure \ref{23}. When $\alpha=0$,
with any value of $\gamma$, we are always back to the same spectrum:
the greybody spectrum (\ref{enfluxU}) for the Unruh vacuum. So one
can compare the other three lines with the black solid line in each
figure to search some features of $\alpha$-vacua. A remarkable
feature of the spectrum is its dependence on $\gamma$. For
$0<|\gamma|\leq\frac{\pi}{2}$, the energy flux is enhanced as
$\alpha$ is tuned up. Between $\frac{\pi}{2}<|\gamma|\leq\pi$, the
flux is depressed with respect to $\alpha$ at low energy.
Particularly, near $|\gamma|=\frac{\pi}{2}$, for small $\alpha$, the
enhancement or depression is invisible in the low energy region.
Remember that as we have mentioned previously, a non-vanishing
$\gamma$ means the breaking of time-reversal symmetry
\cite{Allen85,CM0610}. Unlike eternal black holes, small black holes
in LHC do break the time-reversal symmetry. So it is reasonable to
consider $\alpha$-vacua with $\gamma\neq0$.

Up to now, we have only discussed the spin-0 filed. It is well known
that black holes radiate fields with various spins. $\alpha$-vacua
in de Sitter spacetime for scalar field have been previously
extended to other fields, see \cite{BJM0406,C0410}. If such
extensions go through in Schwarzschild spacetime, their effects
should also be studied to probe $\alpha$-vacua of small black holes
in LHC.

{\bf Acknowledgement}: We would like to thank Miao Li and Wei Song
for useful discussions. We are also grateful to the referees for
valuable comments which have enabled us to improve the manuscript
substantially.

\end{document}